\documentstyle[ltwol]{article}

\def\Journal#1#2#3#4{{#1} {\bf #2}, #3 (#4)}

\def\NPB{{\em Nucl. Phys.} B}
\def\PLB{{\em Phys. Lett.}  B}
\def\PRL{\em Phys. Rev. Lett.}
\def\PRD{{\em Phys. Rev.} D}
\def\ZPC{{\em Z. Phys.} C}

\def\be{\begin{equation}}
\def\ee{\end{equation}}
\def\bea{\begin{eqnarray}}
\def\eea{\end{eqnarray}}

\begin{document}

\title{Review of the PQCD approach to exclusive $B$ decays}

\author{Hsiang-nan Li}

\address{Department of Physics, National Cheng-Kung University, 
Tainan, Taiwan 701, Republic of China}

\address{Physics Division, National Center for Theoretical Sciences,
Hsinchu, Taiwan 300, Republic of China}


\twocolumn[\maketitle\abstracts{ 
I review important aspects of the perturbative QCD approach to exclusive 
$B$ meson decays, concentrating on factorization theorem, gauge 
invariance, end-point singularities, $k_T$ and threshold resummations, 
power counting, penguin enhancement, and CP asymmetries.}]

\section{Introduction}

Perturbative QCD (PQCD) seems to be a successful approach to 
exclusive $B$ meson decays. In this talk I will briefly review 
important aspects of this approach, concentrating on 
factorization of infrared divergences, gauge invariance of
meson light-cone distribution amplitudes, the smearing of end-point 
singularities, Sudakov suppression from $k_T$ and threshold 
resummations, power counting of various topologies of diagrams
for hadronic $B$ meson decays, dynamical enhancement of penguin 
contributions, large strong phases from annihilation diagrams,
and CP asymmetries in $B\to K\pi$ and $\pi\pi$ decays.

\section{Factorization theorem}

We start with the lowest-order diagram for the $B\to\pi$ form factor
in the kinematic region with a fast-recoil pion, which contains a
hard gluon exchanged between the $b$ quark and the spectator quark.
The spectator quark in the $B$ meson, forming a soft cloud around the
heavy $b$ quark, carries momentum of order $\bar\Lambda=M_B-m_b$,
$M_B$ ($m_b$) being the $B$ meson ($b$ quark) mass. The spectator 
quark on the pion side carries momentum of $O(M_B)$ in order to form
the fast-moving pion with the $u$ quark produced in the $b$ quark 
decay. These dramatic different orders of magnitude in momenta 
explain why the hard gluon is necessary. Based on the above argument, 
the hard gluon is off-shell by order of $\bar\Lambda M_B$. As 
explored later, this scale, characterizing heavy-to-light decays, is 
important for developing the PQCD formalism of exclusive $B$ meson decays. 

At higher orders, infinitely many gluon exchanges appear.
If all these gluons are hard, higher-order contributions are 
calculable in perturbation theory. Unfortunately, these diagrams generate
infrared divergences. There are two types of infrared divergences, soft
and collinear. Soft divergences come from the region of a loop momentum
$l$, where all its components vanish:
\begin{eqnarray}
l^\mu=(l^+,l^-,l_T)\sim (\bar\Lambda,\bar\Lambda,\bar\Lambda)\;.
\label{sog1}
\end{eqnarray}
Collinear divergences are associated with a massless quark
of momentum $P\sim (M_B,0,0_T)$. In the collienar
region with $l$ parallel to $P$, the components of $l$ behave like
\begin{eqnarray}
l^\mu\sim (M_B,\bar\Lambda^2/M_B,\Lambda)\;.
\label{sog2}
\end{eqnarray}
In both regions the invariant mass of the radiated gluon diminishes as
$\bar\Lambda^2$, and the corresponding loop integrand may diverge like
$1/\bar\Lambda^4$. As the phase space for loop integration vanishes 
like $d^4 l\sim \bar\Lambda^4$, logarithmic divergences are 
generated.

It has been proved \cite{L4} that the soft divergences in the 
$B\to\pi$ form factor can be factorized into a light-cone $B$ meson 
distribution amplitude, and that the collinear divergences can be 
factorized into a pion distribution amplitude. The remaining finite piece 
is assigned into a hard part. Factorization of infrared divergences is
performed in momentum, spin, and color spaces.  Factorization in 
momentum space means that a hard part does not depend on the 
loop momentum of a soft or collinear gluon, which has been absorbed 
into a meson distribution amplitude. Factorization in spin and 
color spaces means that there are separate fermion and color 
flows between a hard part and a distribution amplitude, respectively. 
To achieve these, we rely on the eikonal approximation for loop 
integrals in leading infrared regions, the insertion of the Fierz 
identity to separate fermion flows, and the Ward identity to sum up 
diagrams with different color structures. Under the eikonal 
approximation, a soft or collinear gluon is detached from the
lines in a hard part and in other distribution amplitudes. The Fierz
identity decomposes a full amplitude into contributions characterized by
different twists. The Ward identity is essential for deriving
factorization theorem in a nonabelian gauge theory. 
 
I emphasize that the hard scale $\bar\Lambda M_B$ is essential for
constructing a gauge invariant $B$ meson distribution amplitude,
\begin{eqnarray}
\phi_B(x)&=&\frac{1}{\sqrt{2N_c}M_B}
\int\frac{dy^-}{2\pi}e^{ixP_B^+y^-}
\langle 0|{\bar q}(y^-)\frac{\gamma_5\not v}{2}
\nonumber\\
& &\times P\exp\left[-ig\int_0^{y^-}dzn\cdot A(zn)\right]
\nonumber\\
& &\times b_v(0)|B(P_B)\rangle\;,
\label{bw}
\end{eqnarray}
with the $B$ meson momentum $P_B$, the velocity $v=P_B/M_B$,
the light-like vector $n=(0,1,0_T)$, the number of colors $N_c$,
the momentum fraction $x$ associated with the spectator quark, and
the rescaled $b$ quark field 
\begin{eqnarray}
b_v(w)=\exp(iM_B v\cdot w)\frac{\not v+I}{2}b(w)\;.
\end{eqnarray}
The above nonlocal matrix element is gauge invariant because of the
presence of the path-ordered Wilson line integral. A careful investigation
shows that the $O(\alpha_s^2)$ diagram with the second gluon attaching 
the hard gluon contributes to this line integral. That is, this diagram 
contains the soft divergence, which is factorized into $\phi_B$. This is 
possible, only when the hard gluon is off-shell by the intermediate scale
$\bar\Lambda M_B$, rather than by $\bar\Lambda^2$ or $M_B^2$.

A hard part is calculable in perturbation theory. A meson distribution 
amplitude, though not calculable, is universal, since it absorbs 
long-distance dynamics, which is insensitive to a specific decay of the 
$b$ quark into light quarks with large energy release. The 
universality of nonperturbative distribution amplitudes is the 
fundamental concept of PQCD factorization theorem. Because of this 
universality, one can extract distribution amplitudes from experimental 
data, and employ them to make model-independent predictions for other 
processes.

\section{End-point singularities}

After developing factorization theorem, we calculate the $B\to\pi$ form 
factor $F^{B\pi}(q^2)$ at large recoil, where $q$ denotes the 
lepton pair momentum. A difficulty immediately occurs. The lowest-order 
diagram for the hard part, which was mentioned above, is proportional to 
$1/(x_1 x_2^2)$, $x_1$ ($x_2$) being the momentum fraction associated 
with the spectator quark on the $B$ meson (pion) side. If the pion 
distribution amplitude vanishes like $x_2$ as $x_2\to 0$ (in the 
leading-twist, {\it i.e.}, twist-2 case), $F^{B\pi}$ is logarithmically
divergent. If the pion distribution amplitude is a constant as $x_2\to 0$ 
(in the next-to-leading-twist, {\it i.e.}, twist-3 case), $F^{B\pi}$ even 
becomes linearly divergent. These end-point singularities have caused 
critiques on the perturbative evaluation of the $B\to\pi$ form factor.

Several methods have been proposed to regulate the above end-point
singularities. An on-shell $b$ quark propagator has been subtracted 
from the hard part as $x_2\to 0$ in \cite{ASY}. However, this
subtraction renders the lepton energy spectrum of the semileptonic
decay $B\to\pi l\bar\nu$ vanishes as the lepton energy is 
equal to half of its maximal value. Obviously, this vanishing is
unphysical. The subtraction also leads to a value of $F^{B\pi}(0)$, 
which is much smaller than the expected one 0.3. A lower bound of 
$x_2$ of $O(\bar\Lambda/M_B)$ has been introduced in \cite{BTF}
to make the convolution integral finite. However, the outcomes 
depend on this cutoff sensitively, and PQCD loses its predictive 
power. 

A QCD-consistent prescription has been proposed in \cite{LY1}, where
parton transverse moemta $k_T$ are retained in internal
particle propagators involved in the hard part. The inclusion
of $k_T$ certainly brings in large double logarithms 
$\alpha_s\ln^2(k_T/M_B)$ through radiative corrections. These large 
logarithms should be resummed in order to improve the perturbative 
calculation. The $k_T$ resummation \cite{CS,BS,LS} then sets
a distribution of $k_T$, such that the average of $k_T^2$ 
is numerically around $\langle k_T^2\rangle\sim \bar\Lambda M_B$
for $M_B\sim 5$ GeV. The off-shellness of internal particles
then remain of $O(\bar\Lambda M_B)$ even in the end-point region,
and the end-point singularities are smeared out. This is so-called 
Sudakov suppression. 

Recently, another type of resummation has been observed. It is not
difficult to verify that the loop correction to the weak decay vertex 
produces the double logarithms $\alpha_s\ln^2 x_2$ \cite{L5}. These
double logarithms can be factored out of the hard part systematically,
and grouped into an exclusive quark jet function. If the end-point
region is important, these large logarithms need to be resummed \cite{KPY}
in order to improve the perturbative calculation. The threshold 
resummation \cite{S0,CT} for the jet function results in Sudakov 
suppression, which decreases faster than any power of $x_2$ as $x_2\to 0$, 
and removes the end-point singularities \cite{L5}. In conclusion,
if the PQCD analysis of the $B\to\pi$ form factor is performed 
self-consistently, there exist no end-point singularities.

The mechanism of Sudakov suppression can be easily understood
by regarding a meson as a color dipole. In the region with
vanishing $k_T$ and $x$, the meson possesses a huge extent in
the transverse and longitudinal directions, respectively. That is,
the meson carries a large color dipole. At fast recoil, this large color 
dipole, strongly scattered, tends to emit real gluons. However, these
emissions are forbidden in an exclusive process with final-state particles 
specified. As a consequence, contributions to the $B\to\pi l\bar\nu$ decay 
from the region with vanishing $k_T$ and $x$ must be highly suppressed.

\section{Power counting}

We then discuss two-body hadronic $B$ meson decays, such as
$B\to PP$. These modes involve 
three scales \cite{CL,YL,CLY}: the $W$ boson mass $M_W$, at which the matching 
conditions of the effective weak Hamiltonian to the full Hamiltonian 
are defined, the typical scale $t\sim \sqrt{\bar\Lambda M_B}$ of a hard
part, which reflects the specific dynamics of a decay mode, and the
factorization scale $\bar\Lambda$, which characterizes infrared 
divergences. Above the factorization scale, perturbation theory is 
reliable, and radiative corrections produce two types of large 
logarithms: $\ln(M_W/t)$ and $\ln(t/\bar\Lambda)$. The former are 
summed by renormalization-group (RG) equations to give the evolution 
from $M_W$ down to $t$ described by the Wilson coefficients $c(t)$, 
while the latter are summed to give the evolution from $t$ to 
$\bar\Lambda$. When the argument of a Wilson 
coefficient $c(\mu)$ is set to $\mu=t$, we have included the logarithmic 
piece of the vertex corrections to the four fermion operators to all
orders. These vertex corrections have been considered in a modified 
factorization approach \cite{Ali,CT98}, which retains the scale 
independence of predictions for hadronic $B$ meson decays. The finite 
piece, being of higher orders in the PQCD approach, is dropped.

The hard part for hadronic $B$ meson decays contains all possible 
Feynman diagrams, such as
factorizable diagrams, where hard gluons attach the valence quarks in
the same meson, and nonfactorizable diagrams, where hard gluons attach
the valence quarks in different mesons. The annihilation topology is also
included, and classified into factorizable or nonfactorizable one
according to the above definitions. Below I will discuss the power counting
of these different topologies of diagrams and argue that all of them 
should be included in the leading-power PQCD analysis.

As explained in Sec.~2, factorizable amplitudes scale like 
$1/\bar\Lambda M_B$, since the end-point singularities do not
exist. For a similar reason, each nonfactorizable diagram also scales 
like $1/\bar\Lambda M_B$. However, because of the soft cancellation 
between a pair of nonfactorizable diagrams, their sum turns out to 
scale like $1/M_B^2$. I emphasize that it is more appropriate to
count the power of each individual diagram, instead of the power 
of sum of diagrams. In some case factorizable contributions
are suppressed by a vanishing Wilson coefficient, such that
nonfactorizable contributions become dominant. For example, factorizable 
internal-$W$ emisson contributions are strongly suppressed by the 
Wilson coefficient $a_2$ in the $B\to J/\psi K^{(*)}$ decays \cite{YL}. 
In some case, such as the $B\to D\pi$ decays, there is no soft 
cancellation between a pair of nonfactorizable diagrams, and 
nonfactorizable contributions also become important \cite{YL}.

A folklore for annihilation contributions is that they are
negligible compared to emission contributions. The annihilation
conributions from the operators $O_{1,2,3,4}$ with the structure
$(V-A)(V-A)$ are small because of helicity suppression. Those
from the operators $O_{5,6}$ with the structure $(S-P)(S+P)$,
though surviving under helicity suppression, are of $O(1/M_B^2)$.
This assumption is reasonable, because the hard
gluon in an annihilaton diagram is off-shell by 
$x_2 x_3M_B^2$, where $x_2\sim O(1)$ and $x_3\sim O(1)$ are
the momentum fractions associated with the two fast outgoing
light mesons. However, this argument applies only to the real 
part of annihilation contributions, but not to the imaginary part. 

To obtain the imaginary part, the internal quark is required to be 
on mass shell. Its propagator, proportional to 
\begin{eqnarray}
\frac{1}{x_2M_B^2-k_T^2}\;,
\end{eqnarray}
then gives 
\begin{eqnarray}
x_2M_B^2=k_T^2\sim \bar\Lambda M_B\;,\;\;\;
x_2\sim \frac{\bar\Lambda}{M_B}\;,
\end{eqnarray}
after considering Sudakov suppression from $k_T$ resummation. This 
smaller $x_2$ implies that the hard gluon 
off-shell only by $\bar\Lambda M_B$ contributes to the imaginary part,
and that the imaginary annihilation amplitudes possess a power behavior 
the same as factorizable emission ones. Note that 
the suppression from threshold resummation decreases the above
estimation by a factor 3 or 4. This is the reason that an 
annihilation amplitude is almost purely imaginary, and its
magnitude is usually few times smaller than the factorizable emission
ones. 

At last, I argue that two-parton twist-3 distribution amplitudes, 
though proportional to the ratio $m_0/M_B$ with the mass $m_0$ 
related to chiral condensate, should be included. As stated before,
the corresponding convolution integral for the $B\to\pi$ form
factor is linearly divergent.
This integral, regulated in some way with an effective cutoff
$x_c\sim \bar\Lambda/M_B$, is proportional to the ratio
$M_B/\bar\Lambda$. Combining the two ratios
$m_0/M_B$ and $M_B/\bar\Lambda$, contributions from
two-parton twist-3 distribution amplitudes are in fact not suppressed
by a power of $1/M_B$:
\begin{eqnarray}
\frac{m_0}{M_B}\int_{\bar\Lambda/M_B}^1\frac{dx_2}{x_2^2}\sim
\frac{m_0}{\bar\Lambda}\;,
\end{eqnarray}
and should be taken into account. Hence, it is appropriate to
claim that the PQCD formalism is complete at leading power.

\section{Phenomenology}

Another important phenomenological consequence related to the special
scale $t\sim\sqrt{\bar\Lambda M_B}$ is the dynamical
enhancement of penguin contributions \cite{KLS}. The RG evolution of 
the Wilson coefficients $C_{4,6}(t)$ dramatically increase as $t<M_B/2$, 
while that of $C_{1,2}(t)$ almost remain constant \cite{REVIEW}. With 
this penguin 
enhancement, the observed branching ratios of the $B\to K\pi$ decays, 
about three times larger than those of the $B\to\pi\pi$ decays,
can be explained for a smaller unitarity angle $\phi_3<90^o$.
Note that the former are dominated by penguin contributions and the 
latter are dominated by tree contributions. In the factorization approach 
\cite{BSW} and in the QCD factorization approach \cite{BBNS}, it is 
assumed that factorizable contributions are not calculable. The leading
contribution to a hadronic decay amplitude is then expressed as a 
convolution of a hard part with a form factor and a meson distribution 
amplitude.  In both approaches the only hard scale is $M_B$ and the 
intermediate scale 
$\bar\Lambda M_B$ can not appear. Therefore, the dynamical enhancement 
of penguin contributions does not exist. It is then difficult to account
for the $B\to K\pi$ data.

To accommodate the $B\to K\pi$ data in the factorization and QCD 
factorization approaches, one relies on the chiral enhancement
by increasng the mass $m_0$ to an unreasonablly large value
$m_0\sim 4$ GeV \cite{WS}. Whether dynamical enhancement or
chiral enhancement is responsible for the large $B\to K\pi$
branching ratios can be tested by measuring the $B\to \phi K$ modes. In 
these modes penguin contributions dominate and the mass $m_0$ is 
replaced by the $\phi$ meson mass $M_\phi\sim 1$ GeV. If the branching 
ratios of the $B\to\phi K$ decays are around $4\times 10^{-6}$
\cite{HMS}, the chiral enhancement may be essential for the $B\to K\pi$
decays. Without including annihilation contributions which depend on
an end-point cutoff sensitively, similar values of the
$B\to\phi K$ branching ratios have been derived in \cite{CY}
using the QCD factorization approach.
If the branching ratios are around $10\times 10^{-6}$ as 
predicted in the PQCD approach \cite{Keum}, the dynamical 
enhancement may be essential.
 
Because of the large imaginary annihilation contributions in the PQCD
formalism, the predicted CP asymmetry ($\sim 30\%$) in the 
$B^0\to\pi^\pm\pi^\mp$ decays \cite{LUY}
dominates over that ($\sim 5\%$) in the QCD factorization approach
and over that ($\sim 10\%$) in the factorization approach \cite{KL}. 
The CP asymmetries in the $B\to K\pi$ decays can also reach 15\% \cite{KLS}.
Future measurements of CP asymmetries can distinguish these different 
approaches to hadronic $B$ meson decays.
For numerical results of all the hadronic modes that have been
studied in PQCD, refer to Dr. Y.Y. Keum's talk in this workshop
\cite{Keum}. 

\section{Conclusion}

In this talk I have briefly reviewed important aspects and
phenomenological consequences of the PQCD approach to exclusive $B$ 
meson decays. I have explained that the end-point singularities do not 
exist in a self-consistent PQCD formalism because of Sudakov suppression 
from $k_T$ and threshold resummations. I have emphasized the special
characteristic scale $\bar\Lambda M_B$, under which a gauge-invaraint
$B$ meson light-cone distribution amplitude can be constructed,
annihilation diagrams should be included in a leading-power analysis,
penguin contributions are dynamically enhanced, and large CP asymmetries 
in the $B\to K\pi$ and $\pi\pi$ decays have been predicted.

More applications of the PQCD approach to other two-body $B$ meson decays
have been discussed in other talks and posters of this workshop.

\section*{Acknowledgments}
I thank I. Bigi, S. Brodsky, H.Y. Cheng, A. Falk, Y.Y. Keum, T. Kuromoto, 
D. London, B. Melic, T. Mannel, T. Morozumi, A.I. Sanda. and L. 
Silvestrini for useful discussions and thank Theory Group of SLAC for 
hospitality during my visit.
The work was supported in part by the National Science Council
of R.O.C. under the Grant No. NSC-89-2112-M-006-033,
and in part by Grant-in Aid for Special Project Research
(Physics of CP Violation) and by Grant-in Aid for Scientific Exchange
from Ministry of Education, Science and Culture of Japan.

\end{document}